\documentclass[runningheads]{llncs}
\usepackage{color}
\usepackage{flushend}
\usepackage{graphicx}
\usepackage{booktabs}
\usepackage[]{xcolor}
\usepackage{tabularx}
\usepackage{multirow}
\usepackage[inline]{enumitem}
\usepackage{xspace}
\usepackage{acronym}
\usepackage{arydshln}
\usepackage{subcaption} 
\usepackage{amssymb}
\usepackage{pifont}
\usepackage[T1]{fontenc}
\usepackage[colorlinks=true,
            linkcolor=blue,
            citecolor=blue,
            filecolor=blue,
            urlcolor=blue]{hyperref}
\usepackage[numbers,sort&compress]{natbib}

\definecolor{midnightgreen}{rgb}{0.0, 0.29, 0.33}

\newcommand{\ikat}{iKAT\xspace}
\newcommand{\cast}{CAsT\xspace}
\newcommand{\llama}{Llama-3\xspace}
\newcommand{\chatgpt}{GPT-3.5\xspace}
\newcommand{\uniquePerTurn}{$\phi$\xspace}

\acrodef{CS}{Conversational Search}
\acrodef{CSA}{Conversational Search Agent}
\acrodef{PTKB}{Personal Text Knowledge Base}
\acrodef{TREC}{TExt Retrieval Conference}
\acrodef{iKAT}{Interactive Knowledge Assistance Track}
\acrodef{CAsT}{Conversational Assistance Track}
\acrodef{NIST}{National Institute of Standards and Technology}
\acrodef{LLM}{Large Language Model}

\acrodef{QPP}{Query Performance Prediction}
\acrodef{IR}{Information Retrieval}
\acrodef{OS}{Open Source}
\acrodef{NLP}{Natural Language Processing}
\acrodef{PEFT}{parameter-efficient fine-tuning}
\acrodef{ICL}{in-context learning}
\acrodef{LoRA}{low-rank adaptation}

\newcommand{\header}[1]{\vspace{1mm}\noindent\textbf{#1}}

\usepackage[T1]{fontenc}
\begin{document}
\title{Improving the Reusability of \\ Conversational Search Test Collections}
\author{} 
\institute{}
\author{Zahra Abbasiantaeb\inst{1}\orcidID{0000-0002-4046-3419} \and
Chuan Meng\inst{1}\orcidID{0000-0002-1434-7596} \and
Leif Azzopardi\inst{2}\orcidID{0000-0002-6900-0557} \and
Mohammad Aliannejadi\inst{1}\orcidID{0000-0002-9447-4172}}
\authorrunning{Z. Abbasiantaeb et al.}
\institute{University of Amsterdam, Amsterdam, The Netherlands \and University of Strathclyde, Glasgow, UK}

\maketitle              %

\begin{abstract}
Incomplete relevance judgments limit the reusability of test collections. 
When new systems are compared to previous systems that contributed to the pool, they often face a disadvantage.
This is due to pockets of unjudged documents (called “holes”) in the test collection that the new systems return.
The very nature of Conversational Search (CS) means that these holes are potentially larger and more problematic when evaluating systems. 
In this paper, we aim to extend CS test collections by employing Large Language Models (LLMs) to fill holes by leveraging existing judgments. 
We explore this problem using TREC \ikat 23 and TREC \cast 22 collections, where information needs are highly dynamic and the responses are much more varied -- leaving bigger holes to fill. 
Our experiments reveal that CS collections show a trend towards less reusability in deeper turns.
Also, fine-tuning the \llama.1 model leads to high agreement with human assessors, while few-shot prompting the \chatgpt results in low agreement with humans. 
Consequently, filling the holes of a new system using \chatgpt leads to higher change in the location of the new system.
While regenerating the assessment pool with few-shot prompting the \chatgpt model and using it for evaluation achieves a high rank correlation with human-assessed pools.
We show that filling the holes using few-shot training the \llama.1 model enables a fairer comparison between the new system and the systems contributed to the pool. Our hole-filling model based on few-shot training of the \llama.1 model can improve the reusability of test collections.

 \vspace{-.3cm}
\end{abstract}

\section{Introduction and Background}
Building reusable test collections in a cost-efficient manner to evaluate current and future systems has been a long-standing challenge in the field of Information Retrieval (IR)~\cite{cleverdon1967,voorhees2005trec}. 
The predominant strategy for creating such collections has been through the use of pooling~\cite{voorhees2005trec,sparck_jones1975} -- where the top-k documents from each contributing system are assessed for relevance.
This approach is considered an acceptable compromise, moving away from the ``\textit{ideal test collection}'' with complete relevance judgments, which is neither feasible nor practical.
While the pooling strategy is fairly robust~\cite{cormack1998efficient_test_collections, buckley2004incomplete_evaluation, sanderson2005ir_evaluation,lu2016effect}, it leads to various evaluation biases (e.g.,~\cite{baillie2007assessed_metric, buckley2004incomplete_evaluation}) where systems that did not contribute to the pool, can be significantly disadvantaged.
This is because documents that haven't been judged are considered irrelevant. 
Therefore, the fewer judged documents returned in a ranking, the lower the retrieval performance ceiling~\cite{baillie2008assessed_metric}.
However, the fewer the judgments required to compare systems the cheaper the test collection. 
Researchers have tried to address these trade-offs by proposing new metrics and methodologies for compensating for the unjudged documents (e.g., ~\cite{buckley2004incomplete_evaluation,aslam2007inferred_ap,baillie2008assessed_metric}), and/or developing new pooling and judgment strategies (e.g., ~\cite{carterette2006minimal_test_collections,lipani2017pooling_strategies,moffat2007strategic_judgments}). Either way, ``holes'' in the pools still remain~\cite{vorhees2022holes,macavaney2023holes}. 
Various studies have proven that the level of missing judgment does not affect the final ranking significantly~\cite{DBLP:conf/sigir/CraswellMYCVS21,DBLP:conf/cikm/Voorhees18}. However, with new advanced ranking models, nowadays, it is more likely that the majority of the top-ranked passages are unjudged.

However, with the advances in the development of powerful \acp{LLM} and other neural-based models, new opportunities have emerged for creating scalable, robust, and reusable test collections at a lower cost. . \acp{LLM} provide the possibility to:
(1) assess large volumes of documents reasonably cheaply, especially compared to human judgments~\cite{Soviero-2024}, 
(2) consistent yet potentially biased evaluations~\cite{Alaofi-llms-fooled-2024}, 
(3) collect judgments under the same conditions at different times, if the \ac{LLM} and prompts are fixed and shared.
(4) typically achieve higher quality than ``typical'' crowd workers~\cite{alaofi2024generative}.

Indeed, recent studies~\cite{meng2024query,khramtsova2024leveraging,thomas2023large,faggioli2023perspectives,rahmani2024syndl,Shivani2024RAG,Shivani2024umbrella} have shown the effectiveness of using \acp{LLM} to generate relevance judgments in the scenario of ad-hoc search automatically. 
These works demonstrate that LLM-based judgments exhibit a high correlation with human judgments. 
Existing works~\cite{thomas2023large,faggioli2023perspectives} prompted commercial \acp{LLM} (e.g., \chatgpt/4) to generate relevance judgments. However, commercial \acp{LLM} come with limitations like non-reproducibility, non-deterministic outputs, and potential data leakage between training and evaluation data, which impend their utility in scientific research~\citep{pradeep2023rankzephyr}.
\citet{macavaney2023holes} and \citet{khramtsova2024leveraging} prompted an \ac{OS} \ac{LLM}, Flan-T5~\citep{chung2024scaling}, for generating relevance judgments. While OS \acp{LLM} are less effective, they do offer the potential for the development of reproducible and reusable test collections at scale. 
This led to efforts by \citet{meng2024query}, who fine-tuned the Llama~\citep{touvron2023llama} model to better condition the \ac{LLM} for performing the task of assigning relevance judgments.
They found that fine-tuned \llama leads to better agreement with human annotators than \chatgpt~\citep{faggioli2023perspectives}. 
Further suggesting that more complete test collections could be produced, with high quality, for lower cost. 
\citet{Shivani2024umbrella} introduced UMbrela, an \ac{OS} toolkit designed for relevance assessment leveraging OpenAI models. UMbrela is designed based on the research by \citet{thomas2023large} and is employed for a large-scale relevance evaluation in the TREC RAG competition \cite{Shivani2024RAG}, demonstrating a strong correlation with human assessors.

While this prospect is very appealing, it is fraught with new, unexplored challenges. For example, \citet{Alaofi-llms-fooled-2024} demonstrated that randomly inserting query words into passages can mislead \acp{LLM}, causing them to incorrectly assess irrelevant passages as relevant.
Of interest, in this work is the notion of grounding. 
Training systems on the judgments of \acp{LLM}, and then evaluating those systems on subsequent test collections, based on judgments from \acp{LLM}, creates a potentially dangerous cycle that may amplify and re-enforce existing biases inherent in \acp{LLM}~\cite{clarke2024llm}. 
Grounding the \ac{LLM}-based judgments given the human judgments provides a mechanism to condition the \acp{LLM} to be more aligned with human annotators -- serving as a means to avoid the hypothesized AI doom loop~\cite{peterson2024ai}. 
To this end, \citet{macavaney2023holes} focus on a setting where the \ac{LLM} is provided with one relevant example in the prompt to help ground the subsequent judgments. 
In this paper, we draw upon this direction in the context of \ac{CS} where we aim to build and augment test collections with grounded \ac{LLM}-based judgments.

\ac{CS} is defined as responding to the user's information needs in the context of the conversation~\citep{radlinski2017cs,azzopardi2018cs,meng2023query,meng2023system,meng2021initiative,meng2020dukenet}. 
In \ac{CS} the user's information need depends on the query, the context of the conversation, and the user's personal preferences~\cite{aliannejadi2024trec}. 
This results in highly dynamic, non-linear conversational trajectories, where a user's information need can be addressed differently depending on the system's interpretation, as those needs evolve and change in response to the presented information.
Therefore, to address such dynamic information needs, systems can follow many different paths. As a result, they retrieve a much larger array of documents, which leads to more unjudged documents. This creates 'bigger holes' in the evaluation of new or future systems, severely limiting the reusability of test collections~\cite{abbasian2024generate}. We propose that \acp{LLM} can effectively and efficiently augment and extend \ac{CS} test collections, thereby improving their reusability.

In this paper, we leverage both commercial and \ac{OS} \acp{LLM} in zero-shot, few-shot, and fine-tuning settings to automatically generate relevance judgments for the \ac{CS} scenario. We compare the generated relevance labels with the official relevance labels by human using various metrics.
Specifically, for commercial \acp{LLM}, we use the \chatgpt model with different prompts, including zero-shot, one-shot, and two-shot prompts.
For \ac{OS} \acp{LLM}, we consider two setups:
(1) directly prompting the \llama.1~\citep{llama3modelcard} model in a zero-shot manner, and
(2) fine-tuning the \llama.1 model on partial official relevance labels from \ac{CS} collections and testing it on the remaining relevance labels.
We evaluate our approach using the TREC \ikat 23~\cite{aliannejadi2024trec} and TREC \cast 22~\cite{owoicho2022trec} \ac{CS} collections, where TREC \ikat 23 serves as a personalized \ac{CS} collection.
In this work, we answer the following research questions:
\begin{enumerate}[label=\textbf{RQ\arabic*}]

    \item How can we measure the reusability of \ac{CS} collections? And how does it change at different conversation depths?
    \label{RQ1} 
    
    \item How do LLM-generated judgments compare to human-generated judgments? \label{RQ2}

     \item Can we make the \ac{CS} collections more reusable using automatic relevance judgments? \label{RQ3}

\end{enumerate}

To answer the RQs, we conduct a set of analyses and experiments. For \ref{RQ1}, we report the number of missing relevant judgments at different conversation depths as a proxy for the collection's reusability.
For \ref{RQ2}, we create a training and test set of relevance labels and compare \llama.1 and \chatgpt in zero-shot, few-shot, and fine-tuning settings. 
We use \chatgpt and \llama.1 to generate relevance labels on the official pool and use it to rank the official TREC runs. We compare the relevance labels generated by \ac{LLM} with human labels in both absolute label prediction and relative ranking of retrieval systems. To answer \ref{RQ3}, we conduct multiple experiments where at each experiment, we remove the unique relevance labels of each run from the pool\footnote{We do not remove the documents that are in common with other runs.}, simulating the case where a new model is being assessed using the original pool. We then generate the relevance labels using \acp{LLM} and use those labels to fill the holes.

\section{Methodology}

\label{sec:methodology}

In this section, first, we define our method to evaluate the reusability of \ac{CS}. Then we explain our approach for using \acp{LLM} as relevance assessors. Lastly, we explain our approach to use \acp{LLM} for enhancing the reusability of \ac{CS} collections. 
To simulate the case where a system needs to be compared against systems in the original pool, we propose the following experimental setups.

\begin{itemize}[leftmargin=*]

    \item \textbf{Leave-one-model-out}: In this scenario, our goal is to compare a new system that did not contribute to the pool against the systems contributed to the pool. To do so, we use the retrieval systems contributed to the pool of the existing collections and remove one system at a time from the pool. We remove the relevance judgment of top-$k$ documents retrieved by a given system ($R$) from the original pool ($P$) to create a modified pool, called $P_{\mathrm{hole}}$. Hence $P_{\mathrm{hole}}$ is a subset of the original pool. Among the top-$k$ documents retrieved by system $R$, we remove the documents that have only been retrieved by system $R$ and did not appear in the list of top-$k$ documents retrieved by other systems.

    \item \textbf{Leave-one-team-out}: The retrieval systems submitted by the same team at TREC are often very similar, so removing just one system while keeping the others from the same team in the pool may not accurately simulate having a significantly different system. To address this, instead of removing a single system's  judgments at a time, we remove all judgments for all systems submitted by a given team. We then follow the same procedure as in the leave-one-model-out scenario to evaluate the impact.
\end{itemize}

\subsection{Conversational reusability}
To establish the extent of the problem (i.e., how big the judgment holes are) while answering \ref{RQ1} on how reusable existing \ac{CS} collections are, we analyze two major \ac{CS} test collections.
One of the simplest ways to estimate the reusability of a test collection is to test whether the basic concept of pooling can be verified. In top-k pooling, the general idea is that the majority of relevant documents appear and get assessed while the pool is created from a sufficiently large and diverse set of retrieval systems. Therefore, as a new system gets evaluated against the pool, the lower the number of unjudged documents is, the more reusable the collection is.
As a proxy to reusability, in this section, we report two metrics per dataset, namely, missing judgments (\uniquePerTurn) and missing relevant judgments ($\phi^+$) for a new system or a set of new systems by one team, that did not contribute to the pooling.

\header{Missing judgments (\uniquePerTurn)} denotes the number of judgment holes for a new system (or a set of systems by one team). These judgment holes can be relevant or irrelevant based on human judgment.

\header{Missing relevant judgments ($\phi^+$)} denotes the number of judgment holes for a new system (or a set of systems by one team) that are relevant based on human judgment. Note that $\phi^+$ is a subset of $\phi$, denoting only the missing judgments of relevant documents. This is particularly important as it demonstrates how much the new system's assessment could be negatively affected by the missing judgments.

\subsection{LLMs as relevance assessors}
\header{Models.} To answer \ref{RQ2} we describe our model setting to use LLMs as relevance assessors. The task of relevance assessment is defined as assigning a relevance score called $s$ to the given passage ($d_i$), given the user's information need and prior conversation. Therefore unlike ad-hoc retrieval, where the information need is captured in a single, self-contained query, \ac{CS} requires understanding the context of the ongoing conversation to assess relevance.
 
Formally, a conversation contains several turns of user-system interactions where user utterance and system response at turn $i$ are indicated by $u_i$ and $r_i$. The context of the conversation at turn $t$ is defined as $c_t$ = \{($u_1$, $t_1$) ... ($u_{t-1}$, $r_{t-1})$\}. The user utterance ($u_t$) is not representative of the user's full information need. Hence, to better understand the information need at turn $i$, the context of the conversation up to this turn is required. Some existing \ac{CS} datasets include the resolved utterance by human, called $u_{t}^{\prime}$, which clarifies co-references and dependencies in the user utterance based on previous turns.  We define function $f$ as follows:
\begin{equation}
\label{eq:relevance-assessment-funct}
s = f(u_{t}^{\prime}, c_t, d_i)~,
\end{equation}
which given the resolved utterance ($u_{t}^{\prime}$), context of the conversation ($c_t$), and document ($d_i$) predicts the relevance score ($s$). The scale of the score $s$ varies from one collection to another collection.  
We use both few-shot prompting and fine-tuning techniques to use \acp{LLM} for relevance assessment. 
\begin{itemize}[leftmargin=*]
    \item {\bf{Few-shot prompting}}: We design three different prompts for few-shot training. Our designed prompts are inspired by the prompts used for relevance assessment in ad-hoc search by \citet{thomas2023large} and \citet{macavaney2023holes}. We use zero-shot, one-shot, and two-shot prompts where in the one-shot one we pass the canonical response as a positive example with maximum relevance score. The canonical response is the perfect system response to the user utterance, provided by the TREC organizers. In our two-shot prompt, we randomly sample one relevant and one irrelevant document from the collection. We design two versions of the one-shot prompt based on using 1) both the context and resolved utterance and 2) only using the resolved utterance.\footnote{We provide our prompts in our GitHub: \url{https://github.com/ZahraAbbasiantaeb/LLMs-for-relevance-judgment-assessment/}.} We use both \ac{OS} and commercial \acp{LLM} and do the relevance assessment task over all query-document pairs from the existing pool. We call the new pool generated by \ac{LLM} as relevance assessors $P_{\mathrm{LLM}}$.

    \item {\bf{Fine-tuning}}: For fine-tuning we split the judgments from the existing collections into train, test, and validation sets. We fine-tune the models on the train set and use the fine-tuned models to assess the relevance of query-document pairs from the test set. In addition, we sample the judgments of few-shot prompting the \acp{LLM} on the test set for comparison with fine-tuned models. We call the new judgment set generated by \acp{LLM} on the test set $P_{\mathrm{LLM}}^{\mathrm{test}}$.

\end{itemize}

\header{\acp{LLM} vs.\ human assessors.} To further answer \ref{RQ2}, we compare the \ac{LLM} and the human judgments.
Given a new judgment set $P_{\mathrm{LLM}}$ (or $P_{\mathrm{LLM}}^{\mathrm{test}}$) generated by LLM, we report the kappa agreement between labels by \acp{LLM} and labels by human on the same set. We report the agreement using both graded labels (0--4) and binary labels. To compute the agreement considering binary labels, we convert scores lower than two into the irrelevant and scores higher than one into the relevant classes. 
In addition, we use the judgments by \ac{LLM}  ($P_{\mathrm{LLM}}$ or $P_{\mathrm{LLM}}^{\mathrm{test}}$) and the judgments by human on the same set to evaluate the retrieval systems. We rank the systems based on the value of nDCG@5 metric and compare two rankings using Kendall's tau ($\tau$) rank correlation metric.

\subsection{LLMs for reusability}
\label{sec:3-llm-for-reuability}
To study the impact of judgment holes on the evaluation of the new retrieval systems and answer \ref{RQ3}, we simulate the leave-one-model-out scenario. 
After removing system $R$ in the leave-on-model-out scenario, we calculate the average number of holes among the top-$k$ documents retrieved by $R$, across all user queries in the dataset. This metric is referred to as \textit{Unjudged@k}. We repeat this process for all systems contributed to the pooling.
Also, after removing system $R$, we fill the holes using \ac{LLM} for relevance assessment and form the new assessment set called $P_{\mathrm{filled}}$. This assessment pool is a mix of judgments by human and \ac{LLM}. We evaluate and rank the systems using both the original pool ($P$) and $P_{\mathrm{filled}}$.
We compute the absolute distance between the position of a the system $R$ in the two ranking lists and and call it $D$. For example, if the system $R$ is ranked 3rd using the original pool ($P$), and ranked 5th using the $P_{\mathrm{filled}}$, then $D=2$.

To study the application of \acp{LLM} to enhance the reusability of existing collections (\ref{RQ3}), we test them on leave-one-team-out scenarios. We repeat the same process mentioned above and form the $P_{\mathrm{filled}}$. After removing all retrieval systems submitted by one team, we fill the holes using judgments by \ac{LLM} and form the new assessment set ($P_{\mathrm{filled}}$). We evaluate and rank all retrieval systems using the original pool ($P$) and $P_{\mathrm{filled}}$ over turns with different depth from the conversation. Then, we compute the rank correlation ($\tau$) and rank change ($D$) using ($P$) and $P_{\mathrm{filled}}$ for ranking the retrieval systems.

\begin{figure}[t]%
    \centering
    \begin{subfigure}[t]{0.43\textwidth}
        \centering
\includegraphics[width=\textwidth]{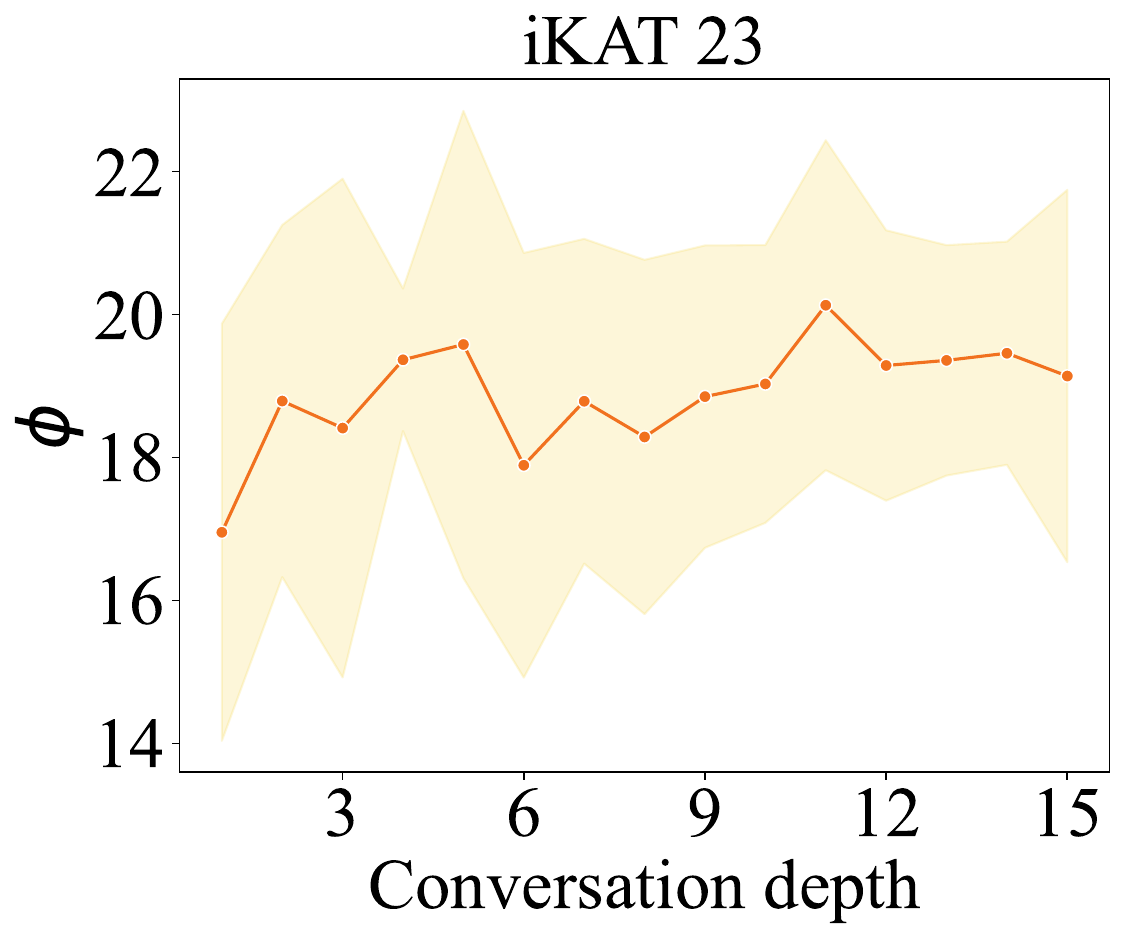}  %
        \label{fig:figure1}
    \end{subfigure}
        \begin{subfigure}[t]{0.43\textwidth}
        \centering
        \includegraphics[width=\textwidth]{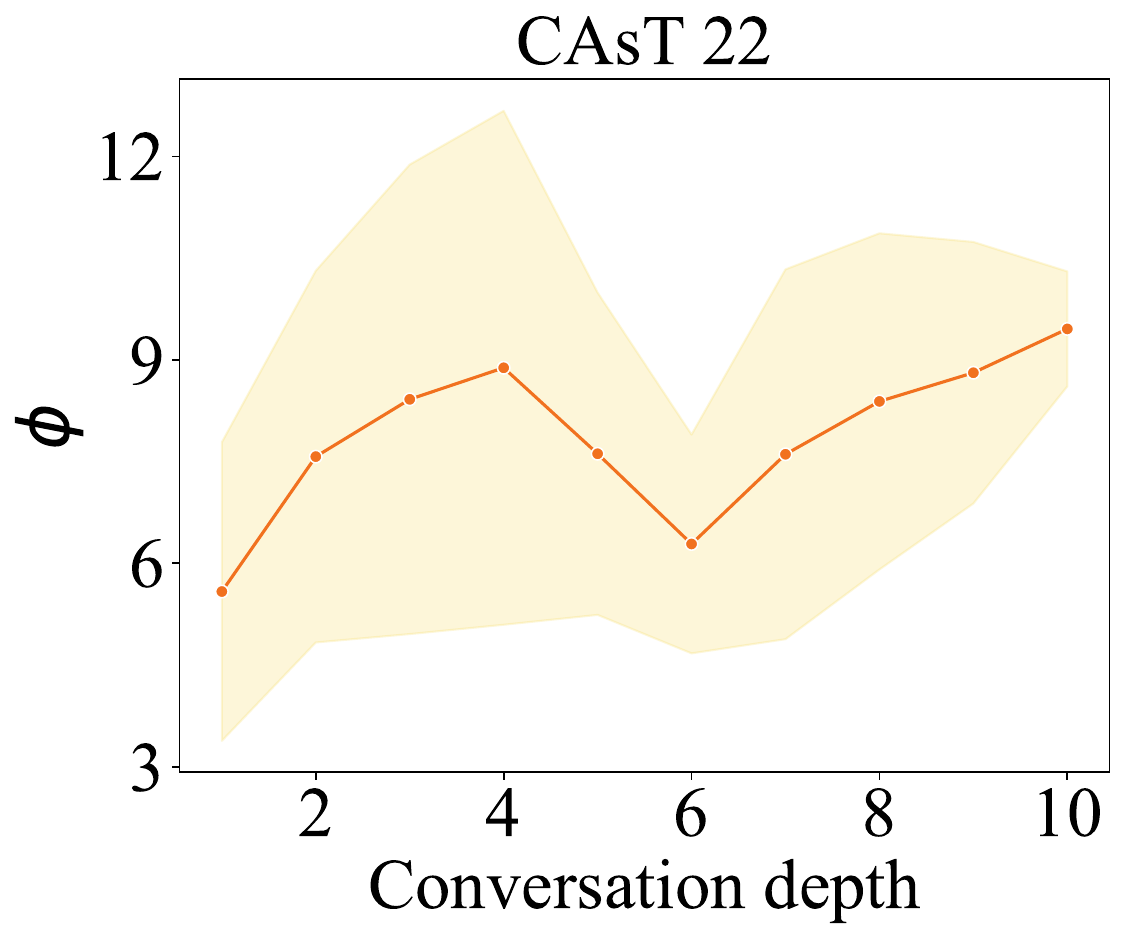}  %
        \label{fig:figure3}
    \end{subfigure}
        \begin{subfigure}[t]{0.43\textwidth}
        \centering
\includegraphics[width=\textwidth]{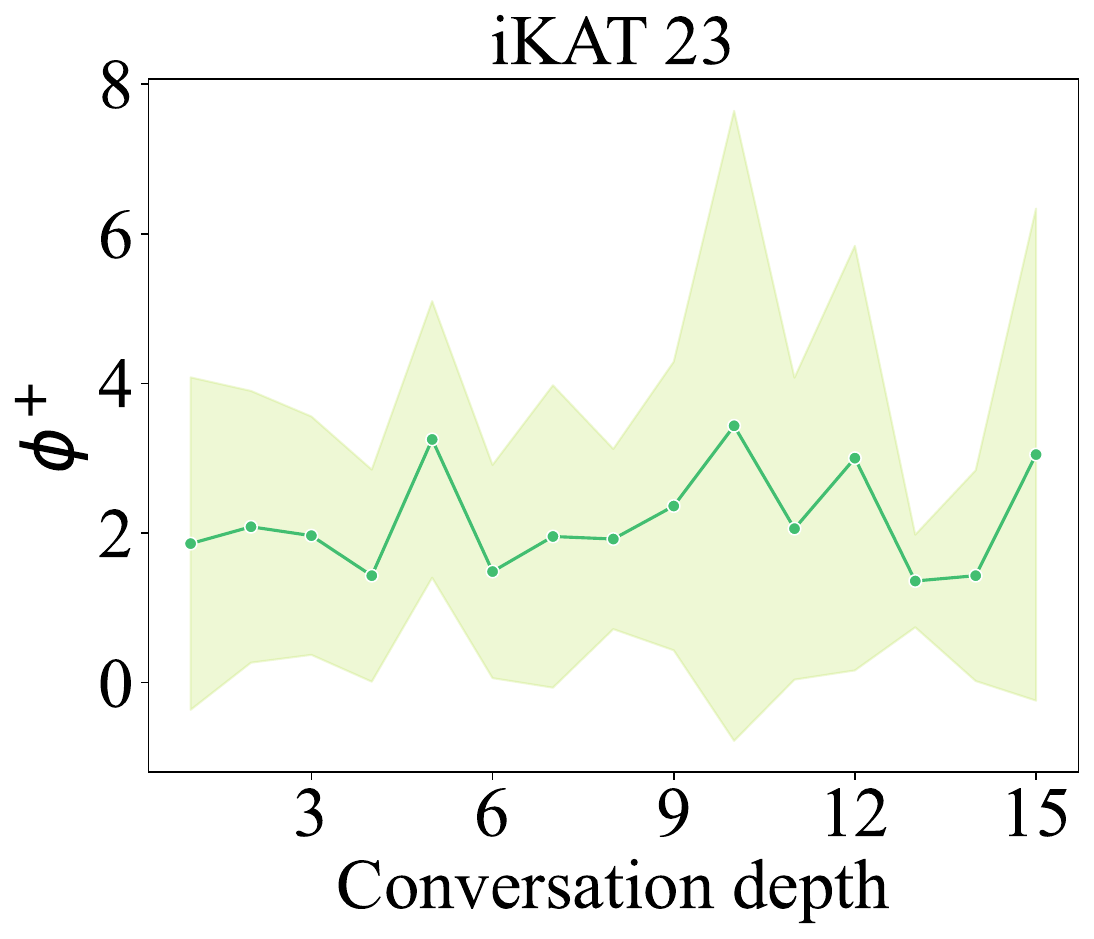}  %
        \label{fig:figure2}
    \end{subfigure}
        \begin{subfigure}[t]{0.43\textwidth}
        \centering
\includegraphics[width=\textwidth]{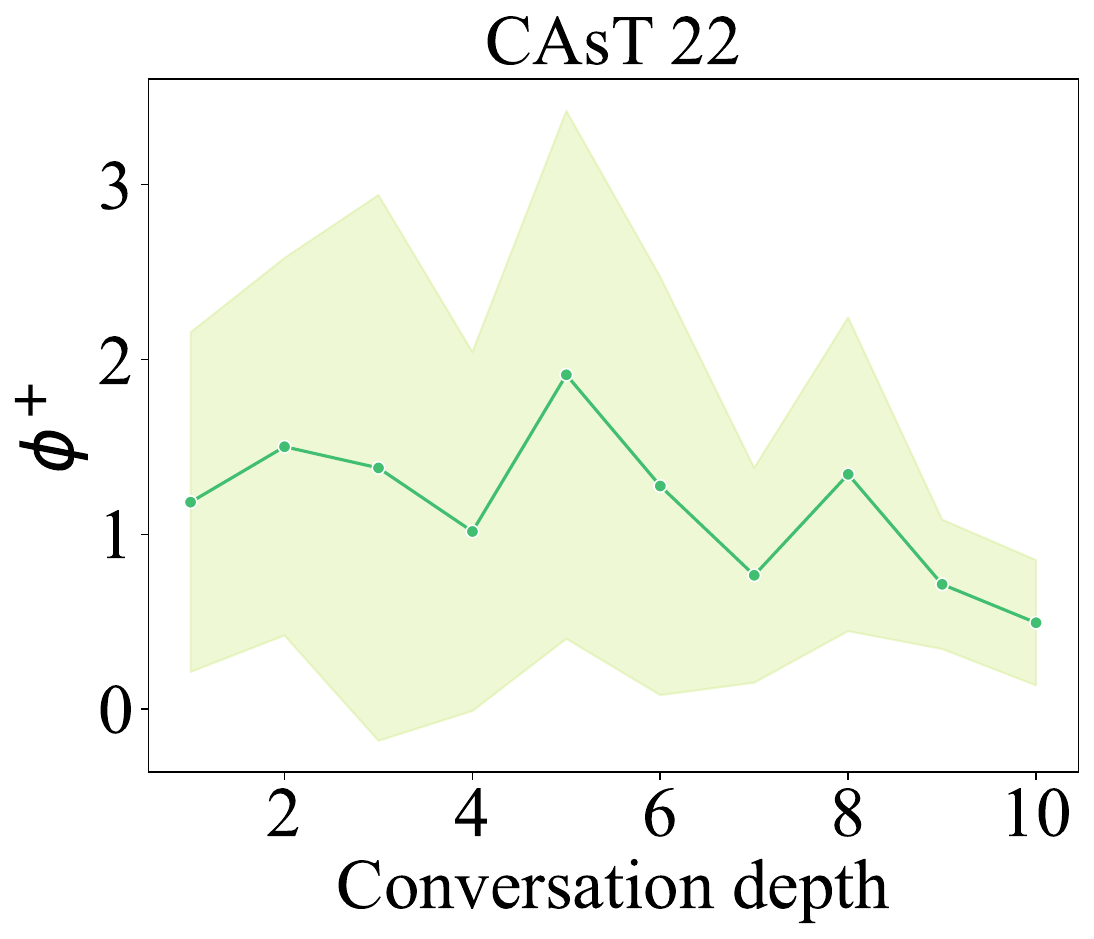}  %
        \label{fig:figure4}
    \end{subfigure}
    \caption{Distribution of average and standard deviation of the count of unique documents ($\phi$) and relevant unique documents ($\phi^{+}$) retrieved by systems per depth of the conversation. These plot are based on leave-one-team-out scenario.}
    \label{fig:unique-per-turn}

\end{figure}

\section{Experimental Setup}
\label{experimental-setup}

\header{The choice of \acp{LLM}.} 
As a commercial proprietary \ac{LLM}, we consider the GPT-3.5 (\texttt{gpt-3.5-turbo-0125}) model with \texttt{temperature=0} and \texttt{top\_p=1}. 
As an \ac{OS} \ac{LLM}, we experiment with \llama.1 (\texttt{llama-3.1-instruct} (8B))~\cite{llama3modelcard} by both few-shot prompting and fine-tuning. For fine-tuning the \llama.1 model, we follow \citet{meng2024query} to use \ac{PEFT} method, 4-bit QLoRA~\citep{dettmers2023qlora}. Also, we fine-tune the Flan-T5 (\texttt{base})~\cite{chung2024scaling} model with a maximum sequence length of 512 and a learning rate of 2e-5. \looseness=-1

\header{Datasets.} We conduct our experiments on the TREC \ikat 23~\cite{aliannejadi2024trec} and TREC CAsT 22~\cite{owoicho2022trec} collections. In these benchmarks, the relevance of each query-document pair is assessed on the scale of 0 to 4. The TREC \cast 22 and \ikat 23 collections include the assessment of 42,196 and 26,159 query-document pairs, respectively. For fine-tuning, we divide the assessments into train, test, and validation sets. First, we randomly remove a portion of irrelevant documents (those with a score $< 2$) from the pool to balance the number of relevant (score $>= 2$) and irrelevant documents per query. Next, we randomly distribute the documents for each query into the train, test, and validation sets, maintaining a 70\%-15\%-15\% split. We ensure that all user queries appear in the train set, with at least one positive and one negative document per query. The size of train, test, and validation sets of TREC \ikat 23 are 3,582, 917, and 709, respectively. For TREC CAsT 22, the train, test, and validation sets include 8,995, 2,287, and 1,827 samples, respectively.

\header{Retrieval systems.} To evaluate the correlation of the generated assessments, we use the baselines and runs submitted to the TREC \ikat 23 and CAsT 22 collections, ensuring that all models are equally treated while being human-assessed. In total, there are 28 and 38 systems submitted to TREC \ikat 23 and TREC CAsT 22, respectively. We utilize the retrieval outputs from these submissions, as provided by the organizers.

\section{Results}
In this section, we report the results of our designed experiments and answer our research questions.

\subsection{Conversational reusability}

\header{Missing judgments.} We simulate the leave-one-team-out setup and plot the distribution of \uniquePerTurn and $\phi^+$ for all retrieval systems over different depths of the conversation in Figure~\ref{fig:unique-per-turn}. As can be seen, by increasing the depth of the conversation, the average of \uniquePerTurn is increasing and it is not consistent over different depths of the conversation. Hence, the pool becomes more diverse by increasing the depth of the conversation in these \ac{CS} collections. 
In addition, the average value of \uniquePerTurn is around 18.55 for \ikat 23 collection while it is around 7.61 for \cast 22 collection. We argue that a lower value for \uniquePerTurn represents the higher reusability of the test collections which can be the result of 1) using a larger number of retrieval systems for top-$k$ pooling, 2) larger value of $k$ in top-$k$ pooling, 3) covering a diverse set of systems in the pooling, or 4) differences in creating the test collections. 
The \ikat 23 dataset is a personalized test collection while \cast is not personalized. The personalized attribute of the \ikat 23 collection affects the assessment and creation of the topics in \ikat 23. Moreover, fewer systems are used for pooling in \ikat 23 compared to \cast 22. 
It is noteworthy that \ikat 23 also features a number of \ac{LLM}-based systems, which could lead to a more diverse set of documents in the pool, hence having a higher $\phi$ value. However, we cannot verify this effect and leave a deeper analysis of it for the future.

\header{Missing relevant judgments.} As for the relevant missing judgments ($\phi^+$), we observe a different trend. By going deeper in the conversation, the number of relevant missing judgments is decreasing. As we go deeper in the conversation the systems are diverging more and number of irrelevant documents retrieved by them is increasing.
As mentioned earlier, the missing relevant judgments are those that actually lead to an unfair assessment of systems. Therefore, even a small number of missing relevant documents is important. We see in the case of \ikat 23 that at depth 10, we even have cases where the missing relevant documents count up to 8 documents, which clearly puts a new system in an unfair comparison with the others. On \cast 22, we see that the overall trend is much lower (hence better) than \ikat 23 with an average missing relevant document count equal or below 2. But also, if we see the std.\ dev.\ values, a high std.\ dev.\ indicates that there are still systems that are unfairly ranked because of the high number of missing relevant documents.

\header{Upshot.}
We answer \ref{RQ1} by concluding that the \ac{CS} collections show a trend towards less reusability in deeper turns. We think that this is because of more complexity of the user utterances and information needs as the conversation progresses, making it less likely for different systems to return a similar set of documents. \looseness=-1

\subsection{LLMs as relevance assessors}

\begin{table}[t]

    \centering
    \caption{Cohen's kappa agreement between human and \ac{LLM} labels on complete set and test set of pools from TREC \ikat 23 and TREC \cast 22 collections. Agreement over binary and graded labels are indicated by (B) and (G), respectively. }
    
    \begin{tabular}{lccccccccccc}
    \toprule
&&& \multicolumn{4}{c}{TREC \ikat 23 } &&  \multicolumn{4}{c}{TREC \cast 22} \\ \cmidrule{4-7} \cmidrule{9-12} 

LLM & Prompt & Context  & \multicolumn{2}{c}{Complete }  & \multicolumn{2}{c}{Test } &&  \multicolumn{2}{c}{Complete }  & \multicolumn{2}{c}{Test }\\  \cmidrule{4-12}  
  & & & B & G & B & G && B & G & B & G \\ 
\midrule

\multirow{4}{*}{\rotatebox{90}{\chatgpt}} &zero-shot & \ding{55} & 20.7 & 4.1 & 48.9 & 11.7  && 21.7  & 8.7 & 37.1 & 10.0  \\  \cdashline{2-12}

&one-shot & \ding{55}& 23.5  & 13.7 & 49.9 & 21.2  && 24.5  & 10.8 & 43.5 & 13.7 \\
&one-shot & \checkmark & \textbf{27.3}  & \textbf{20.2} & 45.3 & 26.8  && \textbf{29.1}  & \textbf{20.4} & 43.0 & 21.5 \\ \cdashline{2-12}

& two-shot & \ding{55} & 21.8 & 15.2 & 45.4 & 18.4  && 28.5  & 18.3 & 42.9 & 19.2 \\  \midrule

\multirow{4}{*}{\rotatebox{90}{\llama.1}}& one-shot & \ding{55}&  2.4  & 2.8 & 2.4 & 1.7 &&  3.2  & 1.9 & 1.5 & 1.3 \\ 
& one-shot &\checkmark & 5.5  & 3.9 & 2.1 & 3.8 &&  4.8  & 2.8 & 4.0 & 2.8 \\ \cdashline{2-12}
& FT & \ding{55} & - & - &  71.8 & 53.0 && - & - &  \textbf{64.9} & \textbf{47.1}  \\ 
& FT & \checkmark & - & - & \textbf{73.3} & \textbf{55.7} && - & - &  62.2 & 45.4   \\ \midrule
FlanT5 & FT & \ding{55} & - & - & 51.7 & 34.7  && - & - & 46.1 & 32.1 \\

\bottomrule
    \end{tabular}    
    \label{tab:kappa}
\end{table}

\begin{table*}[t]
    \centering
    \caption{Comparison between the relative ranking of retrieval systems of TREC \ikat 23 and TREC \cast 22 collections using \ac{LLM}- and human-generated pools.  The relative ranking is compared using Kendall's tau ($\tau$) metric and retrieval systems are ranked based on nDCG@5 metric.}
    
    \begin{tabular}{lccccccc}
\toprule
&&& \multicolumn{2}{c}{TREC \ikat 23} && \multicolumn{2}{c}{TREC \cast 22} \\ \cmidrule{4-5} \cmidrule{7-8}
LLM & Prompt & Context & Complete  & Test  && Complete  & Test  \\ \midrule

\multirow{4}{*}{\rotatebox{90}{\chatgpt}}  &zero-shot  & \ding{55} & 0.852 & {0.778} &&  0.892 &  0.656\\ 

 & one-shot  &  \ding{55} & \textbf{0.862}  & 0.778 && \textbf{0.900} & 0.676\\ 
 & one-shot  & \checkmark & 0.630   & 0.624  && 0.883 & 0.670\\ 
 
 & two-shot  & \ding{55} & 0.825  & 0.746 && 0.886 & 0.650\\ 
\midrule
  
\multirow{4}{*}{\rotatebox{90}{\llama.1}} 
& one-shot  & \ding{55}   & 0.450 & 0.307 && 0.289 &  0.280\\ 
& one-shot  & \checkmark  & 0.550 & 0.471 &&0.710  &  0.616\\  

& FT & \ding{55} & - & 0.841  && - & 0.881\\ 
& FT & \checkmark & - &  0.751 && - & 0.878\\ \midrule
FlanT5 & FT  & \ding{55}  & - &    \textbf{0.889}    && - & \textbf{0.886}\\   
\bottomrule

    \end{tabular}
    \label{tab:1-shot-rank}
\end{table*}

\header{Agreement with human labels.} 
In Table~\ref{tab:kappa}, we report our proposed LLM assessors' agreement with human labels. The experiments reveal that we can improve the agreement by fine-tuning the \llama.1 model. As can be seen, the fine-tuned \llama.1 model achieves the agreement of 71.8 and 64.9 on the binary level over TREC \ikat 23 and TREC \cast 22 collections, respectively.
Moreover, the one-shot learning on the \chatgpt model has the highest agreement with human labels among zero- and two-shot learning techniques. 
In addition, using the conversation context improves the agreement when using \chatgpt over the both collections. However, for the fine-tuned \llama.1 model, using context improves the agreement only on the TREC \ikat 23 collection.

\begin{figure}[t]
    \centering
\includegraphics[width=0.6\textwidth]{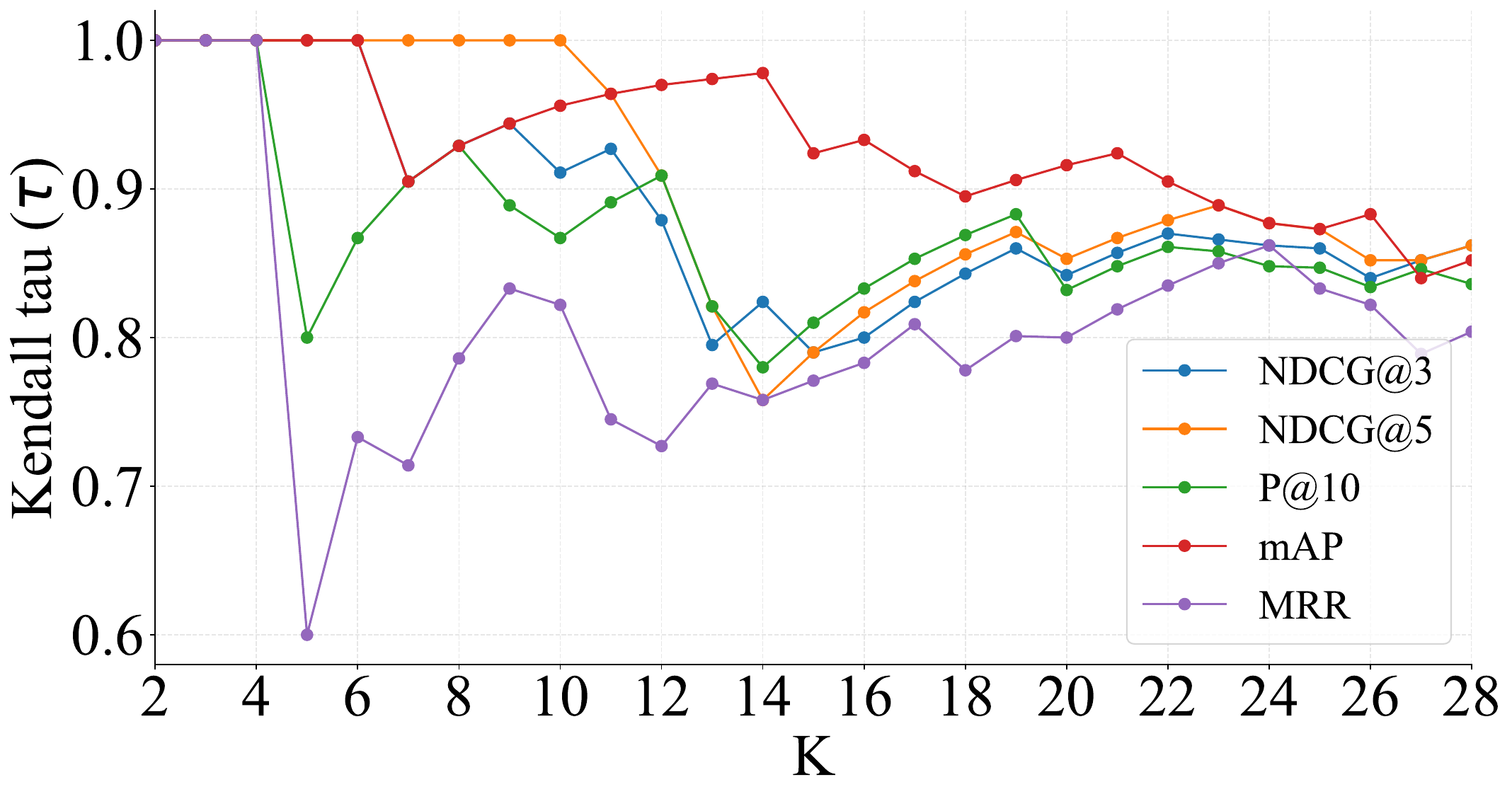}
    \caption{Rank correlation between the $K$ best-performing systems from TREC \ikat 23 using $P$ and $P_{llm}$ pools (based on leave-one-model-out scenario). }
    \label{fig:kendall@k}
\end{figure}

\header{Rank correlation of \acp{LLM} with human.}
We report the correlation between the ranking of retrieval systems using $P_{\mathrm{LLM}}$ and $P$ in Table~\ref{tab:1-shot-rank}. More correlation indicates the effectiveness of the \ac{LLM}-generated pool for evaluating the retrieval systems. 
Surprisingly, the gap between the correlation of the fine-tuned \llama.1 and \chatgpt is highly lower on test data compared to the gap between their agreement with human labels. This could be due to the different labeling biases that the models have where \chatgpt labels could be more different from human labels in terms of absolute numbers (usually more positive~\cite{faggioli2023perspectives}), but when we compare different documents they are more similar relatively.
The FlanT5 model achieves comparable rank correlation compared to \llama .1 while it exhibits a lower agreement on generated labels compared to \llama.1. 
In addition, the \chatgpt model achieves the highest rank correlation using the one-shot prompt on the entire set from TREC \ikat 23 collection. 
Moreover, using the context in the prompt results in a lower rank correlation while it increases the agreement. We conclude that the high agreement between \ac{LLM} labels and human labels does not necessarily result in a higher rank correlation.

\begin{figure}[t]
    \centering
    \begin{subfigure}[t]{0.43\textwidth}
        \centering
        \includegraphics[width=\textwidth]{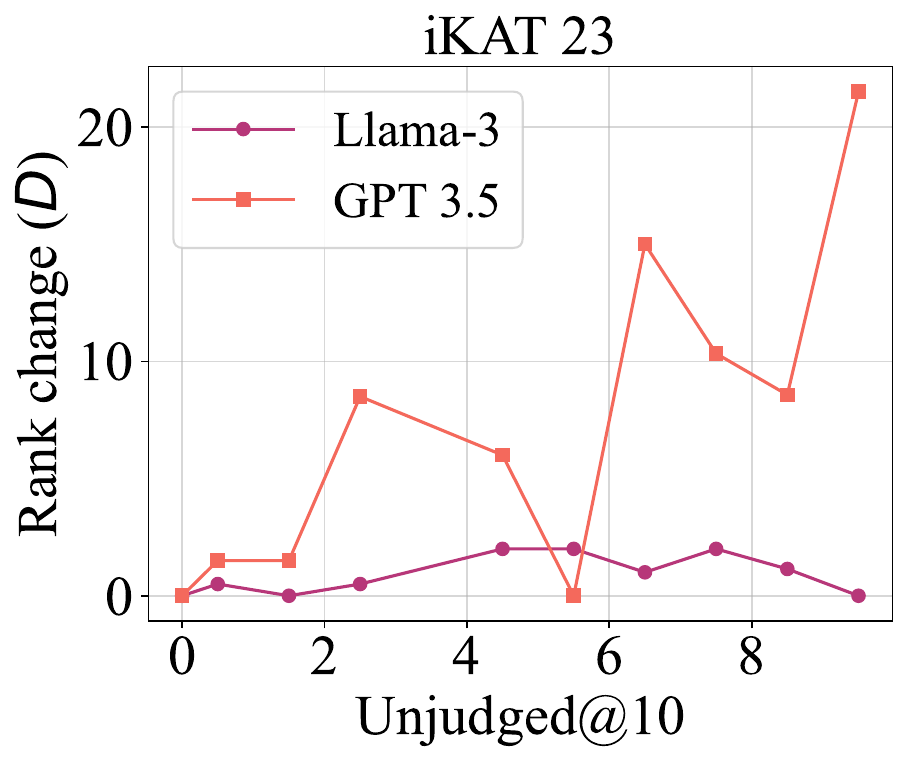}  %
        \label{fig:hole_dist-kat}
    \end{subfigure}
    ~~~
    \begin{subfigure}[t]{0.43\textwidth}
        \centering
        \includegraphics[width=\textwidth]{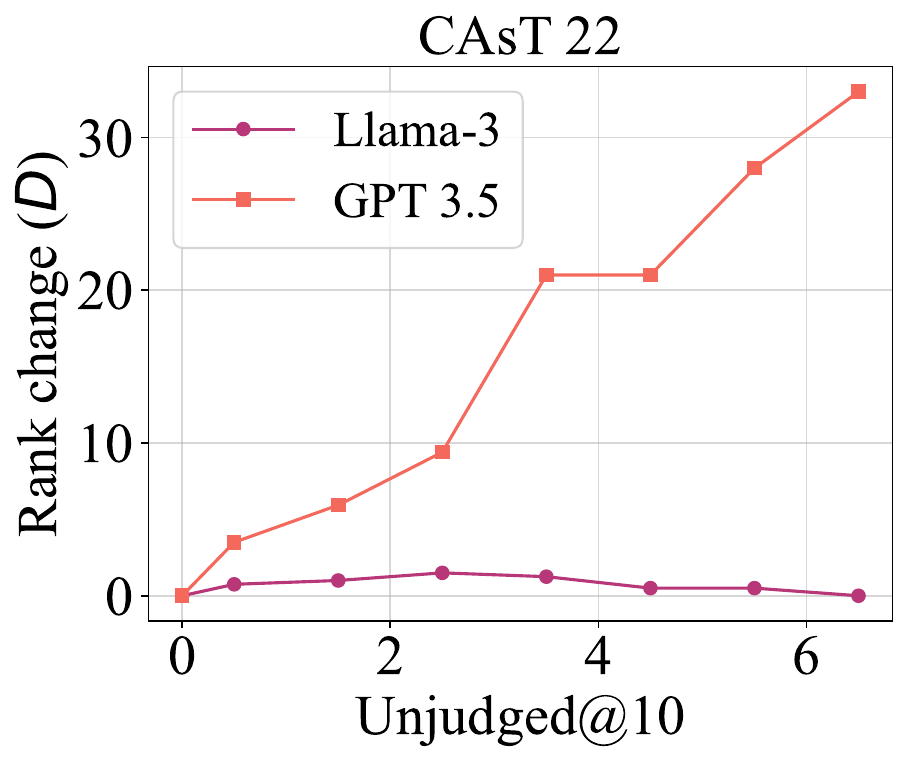}  %
        \label{fig:hole_dist-cast}
    \end{subfigure}
    \caption{Absolute distance between the location of a new system before and after filling the holes using \chatgpt and \llama.1.}
    \label{fig:hole_dist}
\end{figure}

\header{Ranking of top systems.}
In Figure~\ref{fig:kendall@k}, we show the correlation between the relative ranking of $K$ best-performing  systems using $P_{\mathrm{LLM}}$ and $P$.
The relative ranking using $P_{\mathrm{LLM}}$ is the same as using the $P$ over all ranking metrics for four best-performing systems ($K=4$).
 The relative ranking of the top 10 best-performing systems based on nDCG@5 using the judgments by \ac{LLM}, is the same as using human judgments ($\tau=1$). According to Figure~\ref{fig:kendall@k}, as the value of $K$ increases (more systems are included in the comparison), the value of $\tau$ converges. This finding represents the reliability of \ac{LLM}-generated judgments for evaluation. 

\header{Upshot.}
We answer \ref{RQ2}, by concluding that fine-tuning the \acp{LLM} can lead to high agreement for absolute label prediction. However, a low agreement does not result in a low rank correlation when ranking retrieval systems. In addition, by one-shot prompting the \chatgpt we can achieve a high rank correlation.

\subsection{\acp{LLM} for reusability}
\begin{figure}[t]
    \centering
    \begin{subfigure}[t]{0.43\textwidth}
        \centering
        \includegraphics[width=\textwidth]{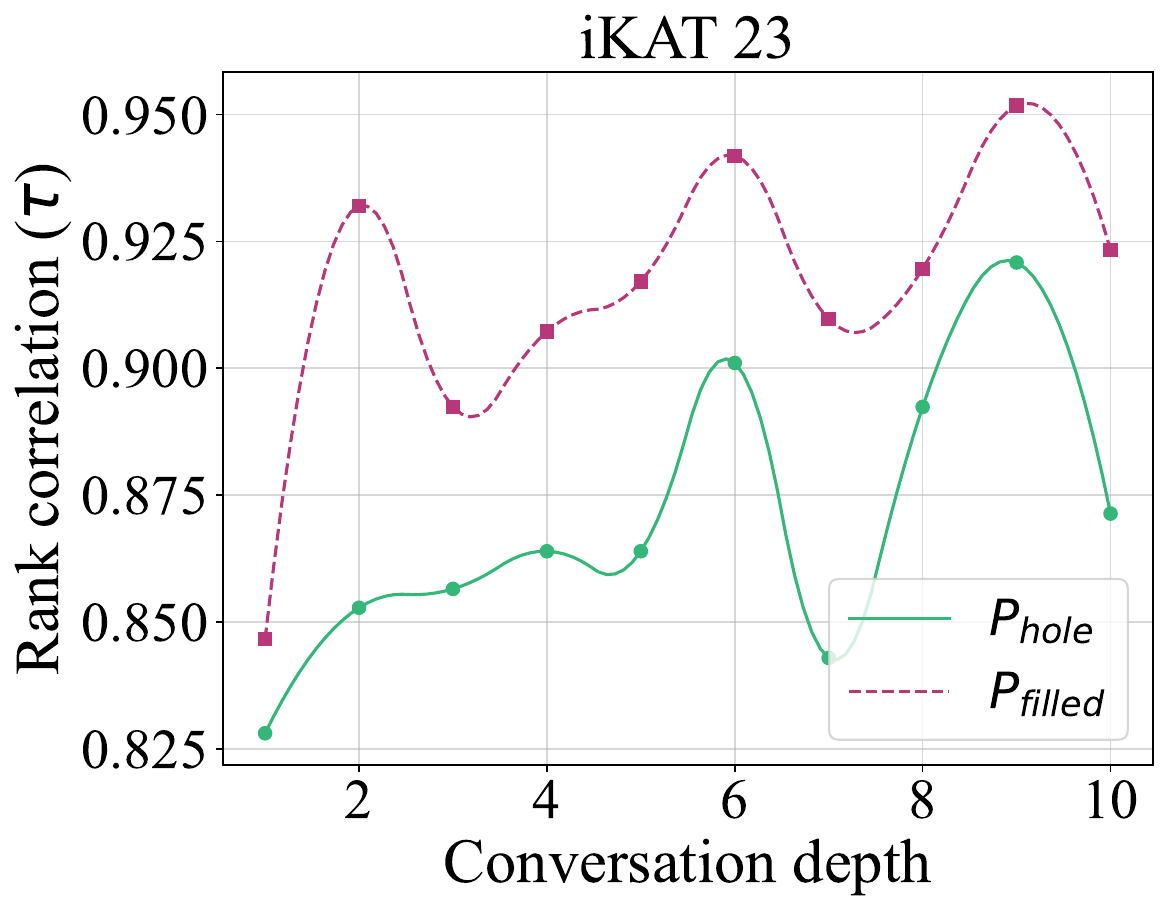}  %
        \label{fig:figure8}
    \end{subfigure}
    \hfill
    \begin{subfigure}[t]{0.43\textwidth}
        \centering
        \includegraphics[width=\textwidth]{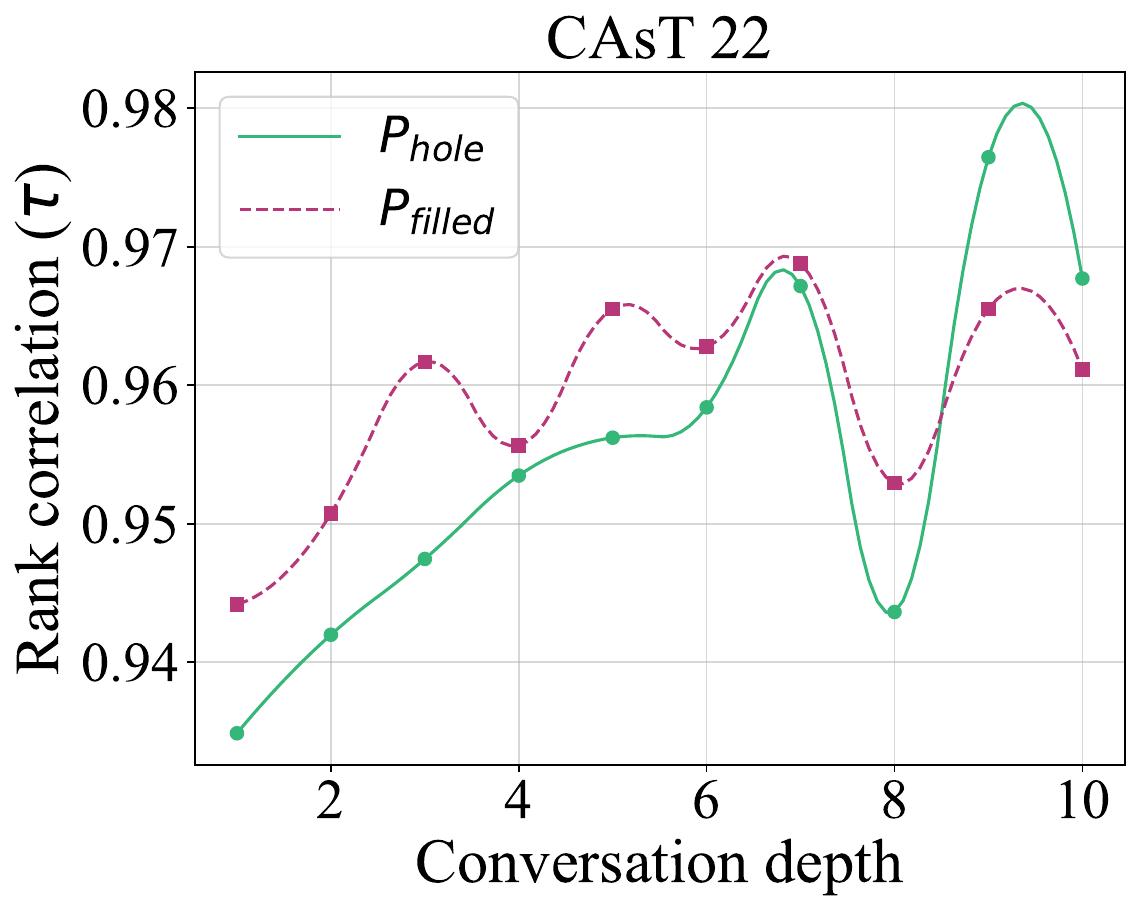}  %
        \label{fig:figure5}
    \end{subfigure}
    \hfill
            \begin{subfigure}[t]{0.43\textwidth}
        \centering
        \includegraphics[width=\textwidth]{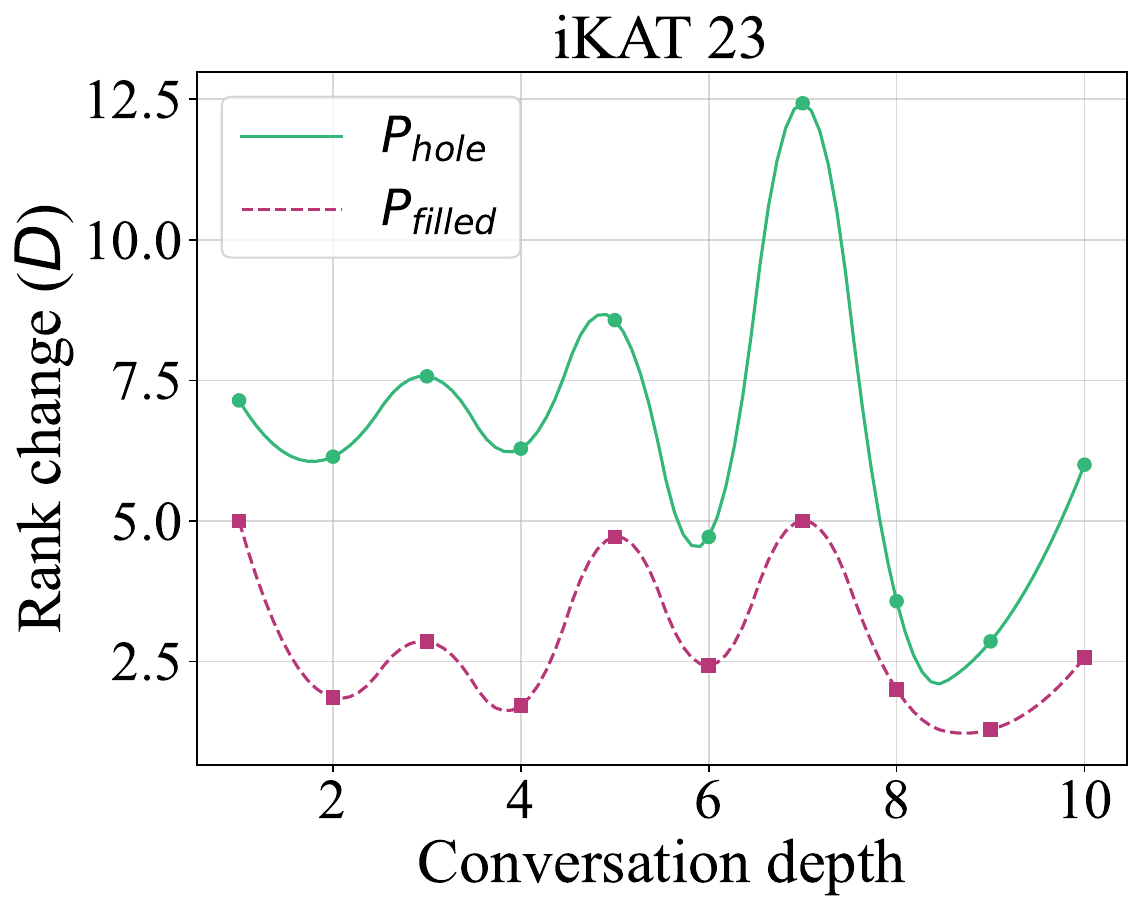}  %
        \label{fig:figure6}
    \end{subfigure}
    \hfill
    \begin{subfigure}[t]{0.43\textwidth}
        \centering
        \includegraphics[width=\textwidth]{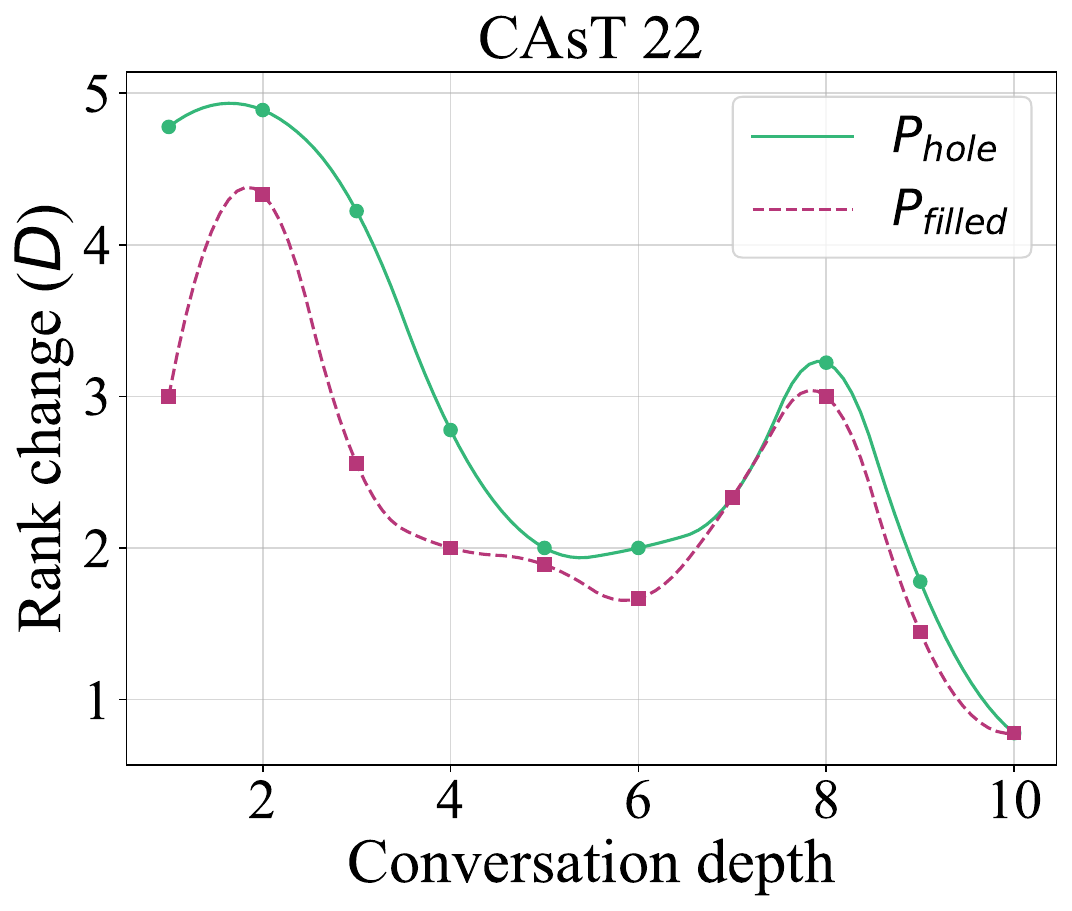}  %
        \label{fig:figure7}
    \end{subfigure}
    \caption{Average rank correlation ($\tau$) of systems and rank change ($D$) of a new system using $P_{\mathrm{filled}}$ or $P_{\mathrm{hole}}$ with original pool $P$ over conversational turns with different depth. The $P_{\mathrm{filled}}$ is formed using the judgments by few-shot prompting the \llama.1 model and leave-one-team-out scenario.}
    \label{fig:leave-one-team-out-tau}
\end{figure}

\header{Reliability of \ac{LLM} labels.} To answer {\ref{RQ3}}, we simulate the leave-one-model-out scenario and use one-shot prompting the \chatgpt model with no context to form $P_{\mathrm{filled}}$. 
The value of the absolute distance ($D$) in the two rankings based on the portion of the \textit{Unjudged@10} documents for the new retrieval system is shown in Figure~\ref{fig:hole_dist}. 
As can be seen, as the value of \textit{Unjudged@10} increases, the absolute distance ($D$) increases. This makes sense as we know the \chatgpt model is biased to rate the documents with higher scores compared to human~\cite{faggioli2023perspectives,meng2024query}. 
We re-create the $P_{\mathrm{filled}}$ by one-shot prompting the \llama.1 model. Interestingly, we see that the results of \llama.1 exhibit a completely different trend where the number of judgment holes does not seem to matter. We see in the plot that \llama.1 can consistently rank the missing systems close to its original ranking and even achieves perfect ranking at some points.

To further answer {\ref{RQ3}}, we do the leave-one-team-out analysis at the different depths of the conversation and use one-shot prompting the \llama.1 model to form the $P_{\mathrm{filled}}$. We plot the average rank correlation between the ranking of systems  and the average absolute distance between the position of the new systems using each of $P_{\mathrm{filled}}$ and $P$ pools in Figure~\ref{fig:leave-one-team-out-tau}. To better study the application of \acp{LLM} for enhancing the reusability of the test collections, we also plot the rank correlation and rank change between $P_{\mathrm{hole}}$ and $P$.

According to Figure~\ref{fig:unique-per-turn}, the average number of unique documents (\uniquePerTurn) retrieved by systems increases by increasing the depth and we have more judgment holes per deeper turns. Consequently, we expect the rank correlation between using $P_{\mathrm{hole}}$ and $P$ to decrease by an increase in the depth.  
However, we observe a different pattern in Figure~\ref{fig:leave-one-team-out-tau}, where the value of $\tau$ increases as we go deeper. Looking at the relevance label of unique retrieved documents in Figure~\ref{fig:unique-per-turn}, we realize that even though the value of \uniquePerTurn is increasing, most of these unique retrieved documents are irrelevant. Also, we observe a drop in the average performance of systems, indicating that the turns become generally ``harder'' and systems fail more.
As the conversation progresses (higher depth), modeling the information need of the user becomes more complex and the retrieval systems diverge more.

According to Figure~\ref{fig:leave-one-team-out-tau}, by filling judgment holes using the few-shot learned \llama.1 model, we can achieve a higher correlation with humans compared to using a pool with hole $P_{\mathrm{hole}}$. Also, as can be seen in Figure~\ref{fig:hole_dist}, using \llama.1 for filling the judgment holes helps to rank the new system closer to its real position in the ranking of systems compared to using the pool with hole $P_{\mathrm{hole}}$.

\header{Upshot.} 
We answer \ref{RQ3} by concluding that given a new ranking model with a lot of missing judgments (a larger value for \textit{Unjudged@10}), it is advisable to recreate the whole pool using one-shot prompting the \chatgpt model, rather than augmenting the existing human-created pool by judgments of the \chatgpt model. Because one-shot \llama.1 aligns better with human labels, it is more effective to use it for assessing the missing judgments, rather than re-assessing the whole pool. As our analyses reveal the \ac{CS} collections become less reusable as the conversations progress (i.e., deeper turns). This is where we can take advantage of \ac{LLM}-generated labels even more.

\vspace{-.2cm}
\section{Conclusion}

Our results show that ranking of the retrieval systems on \ac{CS} test benchmarks using human- and \ac{LLM}-generated are highly correlated, even though the \ac{LLM}-generated and human judgments have a low agreement in binary- and graded-level labels, in line with the findings of \citet{faggioli2023perspectives}. 
In addition, the correlation of different IR metrics converges as we add more retrieval systems to the comparison pool, indicating that \ac{LLM}-generated pools are reliable for comparing different retrieval models in \ac{CS}.
Moreover, we show that in the case of evaluating a new retrieval model, filling missing judgments with a zero-shot \llama.1 model results in less significant shifts in the ranking of the corresponding retrieval model compared to using one-shot \chatgpt, perhaps due to higher agreement of the ratings, and because the one-shot \chatgpt model is biased to predict higher relevance scores~\cite{thomas2023large}.
In addition, we show that in \ac{CS} datasets, the number of judgment holes increases and retrieval systems diverge more as the conversation progresses, making the existing collections less reusable for deeper turns. Our experiments show less reusability of \ikat 23 collection compared to \cast 22 collection. We leave further analysis for the reason behind this finding as future work. 
Interestingly, we find that at further turns in a conversation, the information need becomes more complex and the queries are more difficult since the average performance of models declines.
This fact results in poorer comparisons and less re-usability for the new and more effective retrieval systems in the future -- as the chance of having larger holes for future retrieval systems is greater. 
Our hole-filling approach using few-shot training the \llama.1 model ameliorates this problem and enables a fairer comparison between the systems that contributed to the pool and those that did not. In conclusion, by strategically leveraging \acp{LLM}, we can build more robust, reusable, and cost-effective \ac{CS} test collections that facilitate more equitable comparisons between systems both present and future.
In this study we relied on few-shot prompting the \acp{LLM} for filling the holes in leave-one-model-out scenario. We leave the fine-tuning technique in this scenario as feature work.

\bibliographystyle{splncs04nat}
\bibliography{reference}

\end{document}